# Misfit Strain Accommodation in Epitaxial $ABO_3$ Perovskites: Lattice Rotations and Lattice Modulations

A. Vailionis[1,2], H. Boschker[4], W. Siemons[3], E.P. Houwman[4], D.H.A. Blank[4], G. Rijnders[4], and G. Koster[4]

[1]Geballe Laboratory for Advanced Materials, Stanford University, Stanford, California 94305, USA
[2]Stanford Institute for Materials and Energy Sciences, SLAC National Accelerator Laboratory, 2575 Sand Hill Road, Menlo Park, California 94025, USA
[3]Department of Materials Science and Engineering, University of California, Berkeley, California 94720, USA
[4]MESA+ Institute for Nanotechnology, University of Twente, 7500 AE Enschede, The Netherlands

We present a study of the lattice response to the compressive and tensile biaxial stress in $La_{0.67}Sr_{0.33}MnO_3$ (LSMO) and $SrRuO_3$ (SRO) thin films grown on a variety of single crystal substrates: $SrTiO_3$, $DyScO_3$, $NdGaO_3$ and $(La,Sr)(Al,Ta)O_3$. The results show, that in thin films under misfit strain, both SRO and LSMO lattices, which in bulk form have orthorhombic (SRO) and rhombohedral (LSMO) structures, assume unit cells that are monoclinic under compressive stress and tetragonal under tensile stress. The applied stress effectively modifies the $BO_6$ octahedra rotations, which degree and direction can be controlled by magnitude and sign of the misfit strain. Such lattice distortions change the $B$-O-$B$ bond angles and therefore are expected to affect magnetic and electronic properties of the $ABO_3$ perovskites.

## I. Introduction

The presence of a strong electron-lattice correlations in transition-metal oxides with $ABO_3$ perovskite-type structures imply that the lattice distortions play an important part on physical properties in these materials such as colossal magnetoresistance, ferroelectricity, superconductivity, charge ordering, thermoelectricity, etc. For example, it was suggested that the colossal magnetoresistance in manganese oxide $La_{1-x}X_xMnO_3$ ($X$ = Ca, Sr) compounds cannot be explained by double-exchange mechanism alone and it is largely influenced by strong electron-lattice coupling due to the Jahn-Teller effect, i.e. $MnO_6$ octahedra deformations [1]. In $Ca_{1-x}Sr_xVO_3$ system Inoue *et al.* reported that the one-electron bandwidth $W$ can be controlled without changing the number of electrons and by only varying V-O-V bond angle, i.e. $VO_6$ octahedra rotations [2]. Therefore, the physical properties of the perovskite-type materials are strongly coupled to the shape and rotation of $BO_6$ octahedra.

While in bulk materials the lattice distortions can be varied by applying hydrostatic or chemical pressure, for thin epitaxial $ABO_3$ perovskite films a substrate-induced biaxial stress is an effective tool to modify the electron-lattice coupling. It was shown that ferroelectric thin film properties can be changed by varying the sign and degree of a biaxial strain [3]. First-principles calculations predict that octahedral distortions induced by the epitaxial strain affect magnetic and electronic properties of $SrRuO_3$ thin films [4]. The rotations of $BO_6$ octahedra might induce an "improper ferroelectricity" in short period heteroepitaxial $PbTiO_3/SrTiO_3$ superlattices [5]. Scanning transmission electron microsopy combined with the electron energy loss spectroscopy imaging revealed a close relationship between oxygen octahedra rotations and associated changes in electronic properties at the $BiFeO_3$-$La_{0.7}Sr_{0.3}MnO_3$ thin film interface. [6]. Extensive studies demonstrate that both octahedral deformations and octahedral rotations can be manipulated by a strain which in turn allows to control such properties as ferroelectricity, magnetic anisotropy, metal-insulator transition and superconductivity. $BO_6$ octahedra deformations usually are present through Jahn-Teller distortion (e.g. in $LaMnO_3$) or cation displacement (ferroelectricity in $PbTiO_3$, $BaTiO_3$). Another group of $ABO_3$ perovskite-type materials posses almost rigid octahedra and strain accommodation can be mainly achieved through octahedral rotations/tilts.

The general formula for perovskite unit cell can be written as $ABO_3$, where atom $A$ sits in the center of the unit cell with coordinates (1/2 1/2 1/2) and atom $B$ is located at the unit cell corners (0 0 0). The oxygen is then placed between $B$ atoms at {1/2 0 0} positions. In such arrangement $B$ cation is surrounded by six oxygens forming a corner sharing $BO_6$ octahedra. The simplest perovskite structure is cubic, such as of $SrTiO_3$ (STO) at room temperature, and belongs to space group Pm-3m. By substituting $A$ and $B$ cations, a large number of perovskite-like oxides with different properties can be obtained [7,8]. Generally, the substitutions in such oxides must obey a rule imposed by the ionic radii of the cations which is known as the Goldschmidt tolerance factor [9]:

$$t = \frac{R_A + R_O}{\sqrt{2}(R_B + R_O)}, \quad (1)$$

where $R_A$, $R_B$ and $R_O$ are ionic radii of $A$, $B$ cations and oxygen, respectively. Common perovskite-type compounds usually exhibit tolerance factor of $1.05 > t > 0.78$ [10]. The variation in cation ionic radii induces small deformations or rotations of $BO_6$ octahedra and thus lowers the unit cell symmetry from cubic to tetragonal, orthorhombic, rhombohedral, monoclinic or triclinic.

The changes in the unit cell symmetry resulting from different octahedral rotations have been systematized by Glazer [11] and later expanded by Woodward [12,13]. Octahedral rotations in the perovskite-like unit cell can be described as a combination of rotations about three symmetry axes of the pseudo-cubic unit cell: $[100]_c$, $[010]_c$ and $[001]_c$. The relative magnitudes of the tilts are denoted by letters $a$, $b$ and $c$, e.g. $aab$ means equal rotations around $[100]_c$ and $[010]_c$ axes and different tilt around $[001]_c$ axis. Two adjacent octahedra around one of the $<100>_c$ axes can rotate either in-phase or out-of-phase which is indicated by + or − sign, respectively. Two of the simplest examples are $a^0a^0c^+$ and $a^0a^0c^-$ tilt systems, which are shown in Figure 1. Due to octahedral rotations around single $[001]_c$ axis, the lattice parameters along the $a$ and $b$ axes increase to $\sqrt{2}a_c$. In the $a^0a^0c^-$ system, as octahedra rotate out-of-phase around $[001]_c$, the unit cell along that direction doubles to $2c_c$. No change in $c$-axis lattice parameter will occur if octahedra rotate in-phase. Glazer described 23 tilt systems that are possible in perovskite materials with rigid $BO_6$ octahedra. In general, any of the tilt system can be present in epitaxial perovskite films under misfit stress. It is therefore important to learn how strain of different magnitude and sign affects octahedral rotations in thin films and how these rotations couple to physical properties of the materials.

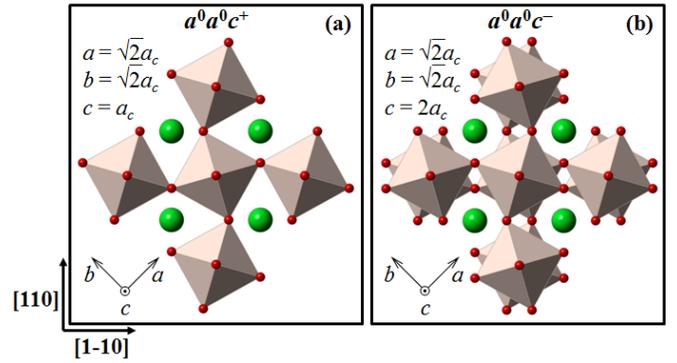

FIG. 1. Schematic representation of $a^0a^0c^+$ (a) and $a^0a^0c^-$ (b) $BO_6$ octahedra rotations in perovskite-type lattice. In both cases the cubic perovskite lattice becomes tetragonal after the rotations take place. The $c$-axis lattice parameter doubles if octahedra assume $a^0a^0c^-$ tilt pattern.

In this work we demonstrate, that epitaxial $SrRuO_3$ and $La_{0.67}Sr_{0.33}MnO_3$ films, structurally behave remarkably similar under misfit strain imposed by the substrate. Under compressive stress both compounds possess (110)-oriented monoclinic unit cell and exhibit $a^+a^-c^-$ tilt system and both switch to a (001)-oriented tetragonal unit cell under tensile stress with tilt system $a^+a^-c^0$. Unit cells of LSMO and SRO show out-of-phase rotations around $[001]_c$ axis for films under compressive stress which are greatly diminished or absent in films under tensile stress. The $BO_6$ octahedra are rotated differently around perpendicular in-plane directions: around $[100]_c$ axis octahedra are rotated in-phase, while around $[010]_c$ the rotations are out-of-phase. Such rotational anisotropy may explain magnetic anisotropy observed in epitaxial LSMO thin films. We believe that such lattice behavior is common to other strained perovskite thin films which in bulk form exhibit orthorhombic or rhombohedral structures: $PrNiO_3$, $LaNiO_3$, $CaRuO_3$, etc.

## II. Experimental

Single TiO$_2$ terminated SrTiO$_3$(001) substrates were obtained with the procedure developed by Koster *et al*. [14]. The treatment of the NdGaO$_3$(110) substrates followed a similar two step procedure with a small modification to the HF solution as described in detail in reference [15]. For the (La,Sr)(Al,Ta)O$_3$(110) and DyScO$_3$(110) substrates no chemical treatment was used. Smooth terraces were obtained after prolonged annealing in a 1 bar oxygen atmosphere, 4 hours at 1050 °C for LSAT [16] and 15 hours at 950 °C for DSO. The epitaxial La$_{0.67}$Sr$_{0.33}$MnO$_3$ [17] and SrRuO$_3$ [18] films were grown by pulsed laser deposition at 800 (700) °C from a stoichiometric targets in an oxygen background pressure of 0.16–0.27 mbar (0.4 mbar, mixed 50% O2+ 50% Ar). A KrF excimer laser ($\lambda$=248 nm) was used with a fluence of 2 J /cm$^2$ and a pulse repetition rate of 5 (4) Hz. The target to substrate distance was fixed at 5 cm. After deposition, the films were cooled down to room temperature at a rate of 10 °C/min in a 1 bar pure oxygen atmosphere. Atomic force microscopy measurements showed smooth surfaces with unit cell high steps.

X-ray diffraction (XRD) measurements were performed using a PANalytical X'Pert materials research diffractometer in high- and medium-resolution modes at the Stanford Nanocharacterization Laboratory, Stanford University as well as at the beamline 7-2 of the Stanford Synchrotron Radiation Laboratory at SLAC, Stanford University. High temperature measurements were performed using an Anton-Paar hot stage. Mn K-edge extended x-ray absorption fine structure (EXAFS) measurements were performed in fluorescence mode at the beamline 4-2 of the Stanford Synchrotron Radiation Laboratory at SLAC, Stanford University. EXAFS data were acquired at room temperature in two orientations: electrical vector of the synchrotron light parallel to the sample surface (in-plane orientation) and with electrical vector at 10 degrees from the sample surface normal (out-of-plane orientation).

## III. Unit cell structures of bulk SRO and LSMO

Room temperature bulk SrRuO$_3$ possesses an orthorhombic crystal structure with space group *Pbnm* (No. 62) and lattice constants $a_o$ = 5.5670 Å, $b_o$ = 5.5304 Å and $c_o$ = 7.8446 Å and it is isostructural with GdFeO$_3$ perovskite [19]. The orthorhombic unit cell is a result of cooperative BO$_6$ octahedra tilts/rotations induced by the mismatch between *A*-O and $\sqrt{2}$(*B*-O) bond lengths. According to Glazer notation, the orthorhombic SRO structure can be described by the tilt system #10: $a^+b^-b^-$ which defines the in-phase octahedral rotations about pseudo-cubic [100]$_c$ axis and mutually equivalent out-of-phase rotations about [010]$_c$ and [001]$_c$ axes. The SRO orthorhombic unit cell can be related to the tilted pseudo-cubic unit cell through the following relationships (see Fig. 2):

$$a_c = \frac{c_o}{2},$$
$$b_c = \frac{\sqrt{a_o^2+b_o^2-2a_ob_o\cos\gamma_o}}{2},$$
$$c_c = \sqrt{\frac{a_o^2+b_o^2-2b_c^2}{2}}, \quad (2)$$
$$\alpha_c = \mathrm{acos}\left(\frac{b_c^2+c_c^2-a_o^2}{2b_cc_c}\right),$$

where $a_c$, $b_c$, $c_c$, $\alpha_c$ and $a_o$, $b_o$, $c_o$, $\gamma_o$ are pseudo-cubic (distorted cubic) unit cell lengths and tilt angle and the distorted orthorhombic unit cell lengths and tilt angle, respectively. Sometimes in the literature the SrRuO$_3$ unit cell is described using approximate pseudo-cubic unit cell:

$$a_p \approx \frac{a_o}{\sqrt{2}} \approx \frac{b_o}{\sqrt{2}} \approx \frac{c_o}{2}. \quad (3)$$

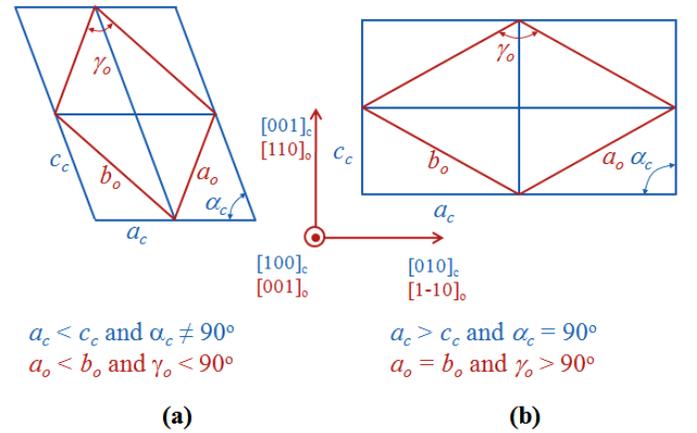

FIG. 2. The relationship between distorted orthorhombic (monoclinic) and pseudo-cubic unit cells: (a) – unit cell under compressive stress and (b) – unit cell under tensile stress.

Bulk La$_{0.67}$Sr$_{0.33}$MnO$_3$ (LSMO) possesses rhombohedral unit cell with space group R3c (No. 167) and lattice constants $a_r$ = 5.471 Å and $\alpha_r$ = 60.43° [20]. In this unit cell octahedra rotations are described by Glazer's tilt system #14: $a^-a^-a^-$ which consists of equivalent out-of-phase rotations about <100> cubic axes. The rhombohedral unit cell can be represented as tilted fcc cubic cell through the following equations [21]:

$$a_f = \frac{\sqrt{2}a_r}{1+\cos\alpha_f},$$
$$\alpha_f = \text{acos}\left(\frac{1-2\cos\alpha_r}{2\cos\alpha_r - 3}\right), \quad (4)$$

where $a_f$, $\alpha_f$, $a_r$, and $\alpha_r$ are the unit cell lengths and angles of tilted fcc cubic and rhombohedral unit cells, respectively. The approximate tilted pseudo-cubic unit cell parameter in this case can be approximated as $a_c \approx a_f/2$.

## IV. Unit cell structures of $ABO_3$ perovskites under tensile and compressive stress

Thin films, that are coherently grown on single crystal substrates, undergo change in the lattice parameters due to mismatch between the unit cells of the growing layer and underlying substrate. According to Frank and van der Merwe [22], the lattice mismatch is defined as $m = (a_l - a_s)/a_s$, where $a_l$ and $a_s$ are the unstrained layer and substrate in-plane lattice constants, respectively. Let us initially describe SRO and LSMO thin film unit cells under compressive and tensile stresses by using a distorted orthorhombic (monoclinic) unit cell with the lattice parameters $a_o$, $b_o$, $c_o$, $\alpha_o$, $\beta_o$, and $\gamma_o$, where $\alpha_o = \beta_o = 90°$ as shown in Figure 2. Such description is perfectly valid for SRO if $\gamma_o = 90°$. LSMO unit cell can also be successfully described as a monoclinic unit cell as has been reported elsewhere [23]. The orientation of such unit cell on (001)-oriented cubic and (110)-oriented orthorhombic/tetragonal single crystal substrates is shown in Figures 3(a) and 3(b), respectively. In this study we used four different substrates: SrTiO$_3$(001) (STO), NdGaO$_3$(110) (NGO), DyScO$_3$(110) (DSO), and (LaAlO$_3$)$_{0.3}$-(Sr$_2$AlTaO$_6$)$_{0.7}$(001) (LSAT). The relationship between pseudo-cubic lattice parameters of the substrates and thin films is illustrated in Figure 3(c).

In order to determine unit cell parameters of strained SRO and LSMO films, we used high-resolution x-ray diffraction. Reciprocal lattice maps (RLM) taken at room temperature using symmetrical and asymmetrical reflections confirm that the SRO and LSMO layers were grown in a fully coherent fashion with respect to the underlying substrate. As an example, RLMs around the (260),(444), (620), and (44-4) reflections of LSMO films under compressive and tensile stresses are shown in Figure 4. Epitaxial SRO thin films exhibit identical behavior [24]. The difference in (260) and (620) atomic plane spacings shown in Fig 4(a) represents a difference in the $a_o$ and $b_o$ film lattice parameters. For thin films under tensile stress (620) and (260) reflections (see Figure 4(b)) show identical positions indicating that $a_o = b_o$. The size and shape of the LSMO and SRO thin film unit cells were determined by refining unit cell parameters using six ($hkl$) reflections: (220), (440), (260), (620), (444), and (44-4). The refined lattice parameters and the calculated strains are listed in Table I. As can be seen from the Table and Figure 2, films under compressive stress possess a unit cell, with $a_o < b_o$, $\alpha_o = \beta_o = 90°$ and $\gamma_o < 90°$, while films under tensile stress exhibit a unit cell with $a_o = b_o$, $\alpha_o = \beta_o = 90°$ and $\gamma_o > 90°$.

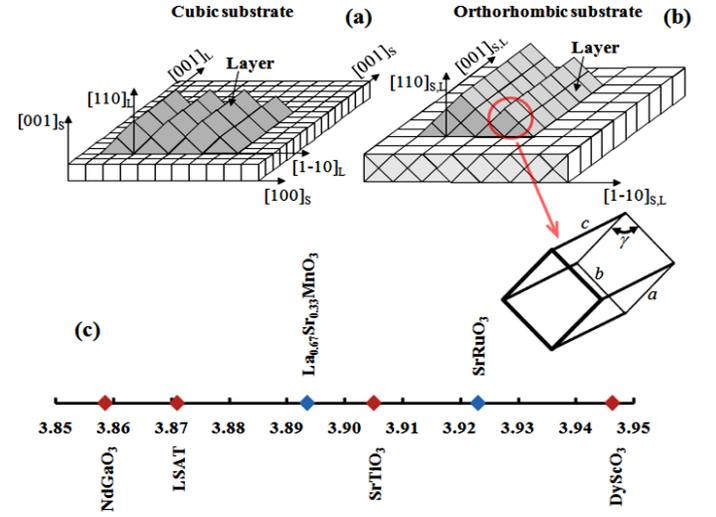

FIG. 3. Schematic representation of the orientation of monoclinic thin film unit cell on (001)-oriented cubic (a) and (110)-oriented orthorhombic (b) substrates. The relationship between pseudo-cubic lattice parameters of SRO and LSMO thin films and the substrates used in this work is shown in (c).

As the stress changes from compressive to tensile, the $ABO_3$ unit cell accommodates that stress differently along perpendicular in-plane directions. Along [001]$_o$ direction stress acts on the length of $c_o$ axis, which is constrained by the lattice parameter of the substrate. In the [1-10]$_o$ direction lattice mismatch is accommodated by the $a_o$ and $b_o$ axes compressive to the tensile, the $b_o$ lattice parameter suddenly decreases and becomes equal to $a_o$ lattice parameter. Films under tensile stress retain the same $a_o$ and $b_o$ lattice parameters which remain almost constant and independent on the magnitude of the tensile strain. The $\gamma_o$-angle, on the other hand, increases continuously from $\gamma_o < 90°$ (compressive stress) to $\gamma_o > 90°$ (tensile stress). Therefore, the tensile stress accommodation along $[1\text{-}10]_o$ direction is achieved by varying only the $\gamma_o$-angle. Due to $b_o$-axis collapse, the orthorhombicity ($b_o/a_o$ ratio) of LSMO and SRO films also abruptly drops to unity as film strain changes from compressive to tensile as shown in Figure 5(b). The volume of the thin film unit cells within the measured strain range increases linearly with strain as predicted by Zayak *et al*. [4] and is shown in Figure 5(c). The absolute volume values of the SRO films are somewhat higher than those calculated, probably due to the underestimate of the lattice constant values in the local spin density approximation [4].

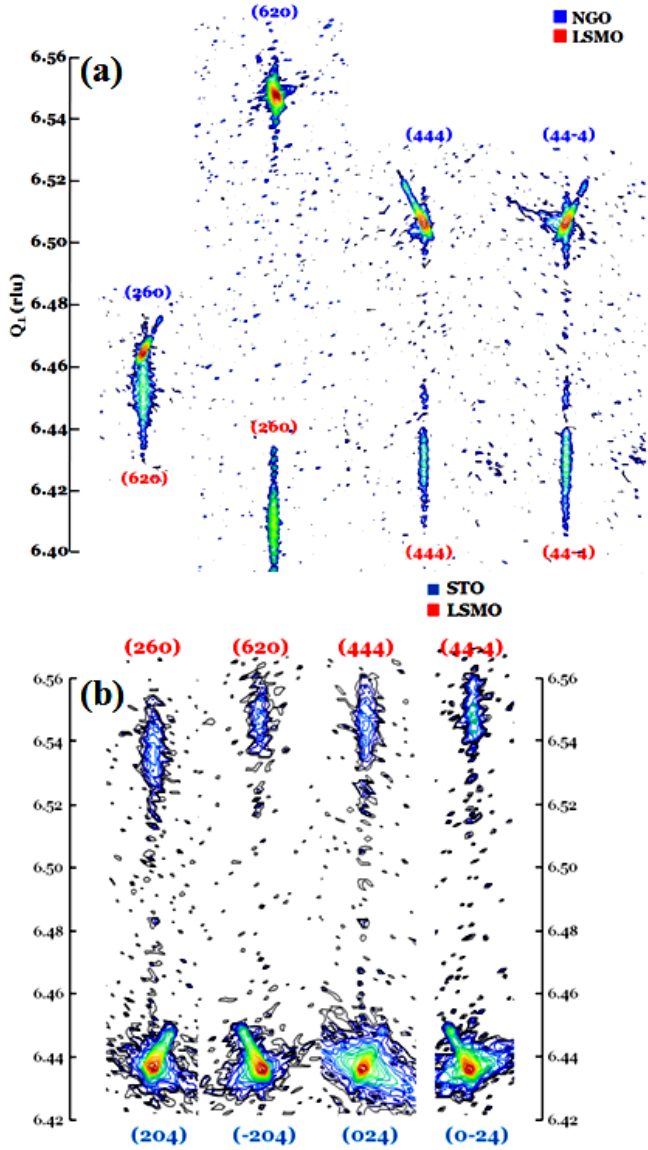

Fig. 4. Reciprocal lattice maps around (620), (260), (444), and (44-4) Bragg reflections of La$_{0.67}$Sr$_{0.33}$MnO$_3$ films grown on NdGaO$_3$(110) substrate under compressive stress (a) and on SrTiO$_3$(001) substrate under tensile stress (b). Here we used $Q_\perp = 4\pi \sin(\theta/\lambda)$, where $\theta$ is the Bragg angle and $\lambda$ =1.540598 Å.

and $\gamma_o$-angle as shown in Figure 2. The $a_o$ and $b_o$ axes are not constrained by the substrate and, together with $\gamma_o$-angle, provide the unit cell with the additional degrees of freedom for strain accommodation. Figure 5(a) shows the behavior of $a_o$, $b_o$ lattice parameters as well as $\gamma_o$-angle as a function of strain along $[1\text{-}10]_o$ direction for LSMO and SRO thin films. The $c_o$-axis (not shown in the Figure), as expected, is strained to the substrate. Along the $[1\text{-}10]_o$ direction strain is acting on a distance $(ab) = \sqrt{a_o^2 + b_o^2 - 2a_o b_o \cos\gamma_o}$ that is matched to the substrate. As the stress changes from

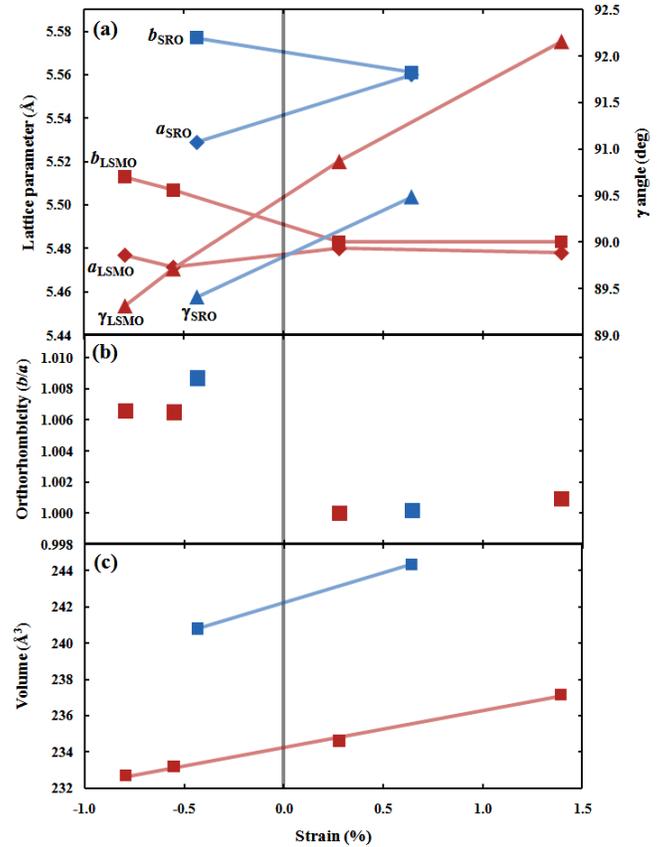

FIG. 5. Variation of some structural parameters of strained La$_{0.67}$Sr$_{0.33}$MnO$_3$ (red) and SrRuO$_3$ (blue) thin films as a function of applied strain: (a) – $a$ and $b$ unit cell lengths and $\gamma$-angle, (b) – orthorhombicity factor ($b/a$), and (c) – volume. Lines are drawn as a guides to the eye. Vertical gray line indicates zero-strain value.

Table I. Refined lattice parameters of SrRuO$_3$ and La$_{0.67}$Sr$_{0.33}$MnO$_3$ thin films grown on different substrates. Bulk values of the orthorhombic SRO and LSMO materials together with the substrate lattice parameters are also shown.

| Substrate & bulk SRO and LSMO | | | | | Strained layer | | | | | |
|---|---|---|---|---|---|---|---|---|---|---|
| | a (Å) | b (Å) | ab (Å) | c (Å) | | a (Å) | b (Å) | ab (Å) | c (Å) | γ (°) |
| NdGaO$_3$ [25] | 5.428 | 5.498 | 7.726 | 7.708 | LSMO/NGO | 5.477 | 5.513 | 7.725 | 7.707 | 89.32 |
| LSAT [26] | 5.476 | 5.476 | 7.744 | 7.740 | LSMO/LSAT | 5.471 | 5.507 | 7.744 | 7.740 | 89.72 |
| SrTiO$_3$ | 3.905 | | | | LSMO/STO | 5.480 | 5.483 | 7.809 | 7.809 | 90.87 |
| DyScO$_3$ [27] | 5.442 | 5.720 | 7.895 | 7.890 | LSMO/DSO | 5.478 | 5.483 | 7.895 | 7.902 | 92.16 |
| LSMO (O) [28] | 5.488 | 5.524 | 7.787 | 7.787 | SRO/STO | 5.529 | 5.577 | 7.813 | 7.810 | 89.41 |
| SRO (O) [19] | 5.530 | 5.567 | 7.847 | 7.845 | SRO/DSO | 5.560 | 5.561 | 7.897 | 7.903 | 90.49 |

| LSMO/NGO | LSMO/LSAT | LSMO/STO | LSMO/DSO | SRO/STO | SRO/DSO |
|---|---|---|---|---|---|
| Strain (%) | | | | | |
| along ab = -0.80  along c = -1.03 | along ab = -0.55  along c = -0.60 | along ab = 0.28  along c = 0.29 | along ab = 1.39  along c = 1.48 | along ab = -0.44  along c = -0.45 | along ab = 0.64  along c = 0.74 |
| Glazer tilt system | | | | | |
| $a^+b^-c^-$ | $a^+a^-c^-$ | $a^+a^-c^0$ | $a^+a^-c^0$ | $a^+a^-c^-$ | $a^+a^-c^0$ |

## V. $BO_6$ octahedra rotations under compressive stress

The change from $a_o < b_o$ for compressive stress to $a_o = b_o$ for the tensile stress unequivocally indicates a change in the $BO_6$ octahedra rotation pattern. We will use Glazer notations to describe the $BO_6$ octahedra rotations and corresponding lattice symmetries of thin epitaxial SRO and LSMO films. Since the layers are coherently strained to the substrate, the film's pseudo-cubic in-plane lattice constants should match the ones of the substrate. Assuming that the substrate's in-plane lattice parameters are equal ($a_s = b_s$), for compressively strained film we have a relationship: $a_c = b_c < c_c$. Moreover, since the $a_o$ and $b_o$ lattice constants of the films are not equal, the pseudo-cubic lattice will be tilted from [001]$_c$ axis by angle $\alpha_c \neq 90°$ (see Fig. 2). According to the Glazer notation, only one tilt system satisfies all these conditions: #9 ($a^+a^-c^-$) [11]. Therefore, we can infer that our SRO and LSMO thin films under compressive stress possess a monoclinic unit cell with space group P2$_1$/m (No. 11) with $a_m = \sqrt{a_c^2 + c_c^2 - 2a_c c_c \cos\alpha_c}$, $b_m = \sqrt{a_c^2 + c_c^2 + 2a_c c_c \cos\alpha_c}$, $c_m = 2a_c$, $\alpha_m = \beta_m = 90°$ and $\gamma_m < 90°$ which is in perfect agreement with our observations shown in Table I.

It is important to note, that the NGO(110) substrate has in-plane lattice parameters that are slightly different and therefore the condition $a_c = b_c$ is not perfectly valid here. In this case, where $a_c \neq b_c < c_c$, a Glazer tilt system #8 ($a^+b^-c^-$) might be more appropriate. Both #9 and #8 tilt systems are very similar. The thin film unit cells under these tilt systems are both monoclinic with the same space group P2$_1$/m (No. 11). The rotation pattern is also preserved under both tilt systems except that in tilt system #8 the $BO_6$ octahedra will be rotated with slightly different magnitudes around [100]$_c$ and [010]$_c$ directions.

The schematic view of the $BO_6$ octahedra rotations under compressive stress is shown in Figure 6(a) and (b). Octahedral rotations in thin films under compressive stress have different patterns around the orthogonal in-plane directions: around the [100]$_c$ direction rotations are in-phase, while rotations around the [010]$_c$ direction are out-of-phase. The $BO_6$ octahedra are also rotated out-of-phase around the [001]$_c$ direction.

## VI. $BO_6$ octahedra rotations under tensile stress

For the tensile stress we have a situation where $a_c = b_c > c_c$ and $\alpha_c = 90°$. In order to allow pseudo-cubic $a_c$ and $b_c$ axes to become longer with respect to the $c_c$-axis, octahedral rotations around the [001]$_c$ direction have to be significantly reduced or absent. The XRD results show that for films under tensile stress, the $a_o$ and $b_o$ lattice constants exhibit similar values [29]. We can therefore conclude that, during the transition from compressive to tensile strain, the rotations of $BO_6$ octahedra around $c_c$-axis are diminished while rotations around $a_c$ and $b_c$ axes are still maintained. Such rotation pattern can be expressed by a tilt system #18 ($a^+a^-c^0$), in which $BO_6$ rotations around $a_c$ and $b_c$ axes are analogous to those under compressive stress, but are absent around $c_c$-axis. This tilt system results in a tetragonal unit cell with the lattice parameters: $a_t = b_t = 2a_c$, $c_t = 2c_c$ and $\alpha_t = \beta_t = \gamma_t = 90°$ and is described by the space group Cmcm (No. 63).

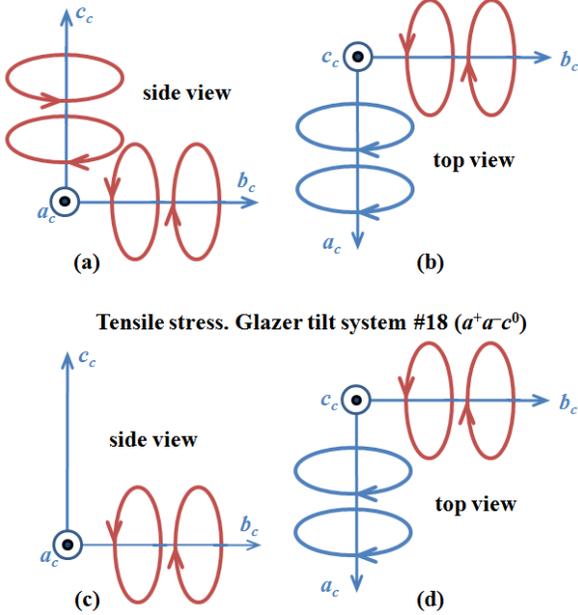

FIG. 6. Schematic illustrations of $BO_6$ octahedra rotations under compressive stress (a) and (b) and under tensile stress (c) and (d). Drawings (a) and (c) are the side view of the unit cell and (b) and (d) are the top view with only in-plane rotations visible.

The Glazer tilt system #16 ($a^+a^+c^0$) also satisfies the condition of a film under tensile stress, where $a_c = b_c > c_c$ and $\alpha_c = 90°$, and will also result in a tetragonal unit cell but with the space group I4/mmm (No. 139). In contrast to the tilt system #18 ($a^+a^-c^0$), $BO_6$ octahedra under the tilt system #16 ($a^+a^+c^0$) are rotated in-phase around both $[100]_c$ and $[010]_c$ in-plane directions. According to Woodward, the former tilt system, #18 ($a^+a^-c^0$), cannot be achieved without some deformations of $BO_6$ octahedra [12]. In order to retain the connectivity of octahedra in this tilt system, either the variation of octahedra angles from 90 degrees by about 0.3° for 1/6 of the bonds or the deviation of $B - O$ distance by about 0.002 Å for 1/3 of the bonds has to be considered. Both deformations are very small and cannot be observed neither by XRD nor EXAFS techniques. Since the tilt system #18 ($a^+a^-c^0$) involves such small deformations, we believe that this tilt system is more likely to occur in thin films under tensile stress than the tilt system #16 ($a^+a^+c^0$). The tilt system #18 ($a^+a^-c^0$) preserves the in-plane octahedral rotation pattern as film goes from compressive to a tensile stress while the Glazer system #16 ($a^+a^+c^0$) requires a change in the rotation pattern of $BO_6$ octahedra which might not be energetically favorable. The schematic view of the $BO_6$ octahedra rotations under tensile stress is shown in Figure 6(c) and (d). Octahedral rotations in thin films under tensile stress have different pattern around orthogonal in-plane directions: around $[100]_c$ direction rotations are in-phase, while rotations around $[010]_c$ direction are out-of-phase. The rotations of the $BO_6$ octahedra around $[001]_c$ direction are greatly diminished or absent. The degree of octahedral rotations around $[100]_c$ and $[010]_c$ directions depends on the magnitude of a mismatch. No rotations should occur in coherently strained film if the substrate in-plane lattice constant exceeds film's B-O-B distance. In that case the B-O-B bond angle will set at 180 degrees and further strain accommodation most likely will be achieved by octahedral deformation. This will result in shorter out-of-plane B-O bond lengths as compared to the in-plane ones. So far we have not been able to determine with sufficient accuracy the Ru-O (in $SrRuO_3$) and Mn-O (in $La_{0.67}Sr_{0.33}MnO_3$) in-plane and out-of-plane bond lengths and therefore cannot verify the deformations of $RuO_6$ and $MnO_6$ octahedra. The effect is still under investigation.

## VII. EXAFS measurements

The tilts of $BO_6$ octahedra were also confirmed by the extended x-ray absorption fine structure. We used linearly polarized synchrotron radiation, which allowed us to probe in-plane and out-of-plane interatomic bonds independently. The R-space Fourier transformed experimental Mn K-edge EXAFS spectra of the LSMO thin films grown on LSAT (compressive strain) and $SrTiO_3$ (tensile strain) substrates are shown in Figure 7. Analysis of the paths contributing to the intensity of each peak shows that the first peak at ~1.5 Å originates from the Mn-O bond. Peak at ~3.4 Å mainly includes contributions of single, double, triple and quadrilateral scattering paths from Mn-Mn, Mn-O-Mn bonds. While for the first peak only the Mn-O single scattering contribution is important, the latter peak intensity is largely affected by multiple scattering contributions. Especially it is very sensitive to the Mn-O-Mn bond angle due to the collinear or non collinear bond arrangement [30,31]. The intensity of the EXAFS peak at ~3.4 Å will be significantly enhanced if Mn-O-Mn bond is collinear (i.e. makes 180 degree angle). On the other hand, the peak intensity will decrease as the bond angle is reduced. As can be seen from Figure 7(a) (in-plane bonds), film under compressive stress (LSMO/LSAT) exhibit ~3.4 Å peak much smaller in intensity than that for film under tensile stress (LSMO/STO). This indicates that the in-plane Mn-O-Mn bonds are more buckled in films under compressive stress due to an additional $MnO_6$ octahedra rotations around $c_c$-axis. The opposite but smaller effect can be seen in Figure 7(b) which shows EXAFS data originating from the out-of-plane bonds. In this case film under compressive stress has straighter out-of-plane Mn-O-Mn bonds than film under tensile stress. EXAFS peak intensities along out-of-plane direction are mainly influenced by $MnO_6$ octahedra rotations around $[100]_c$ and $[010]_c$ axes. It is important to note, that Mn-O bond lengths are roughly the same in all cases supporting a rigid octahedra scenario [32].

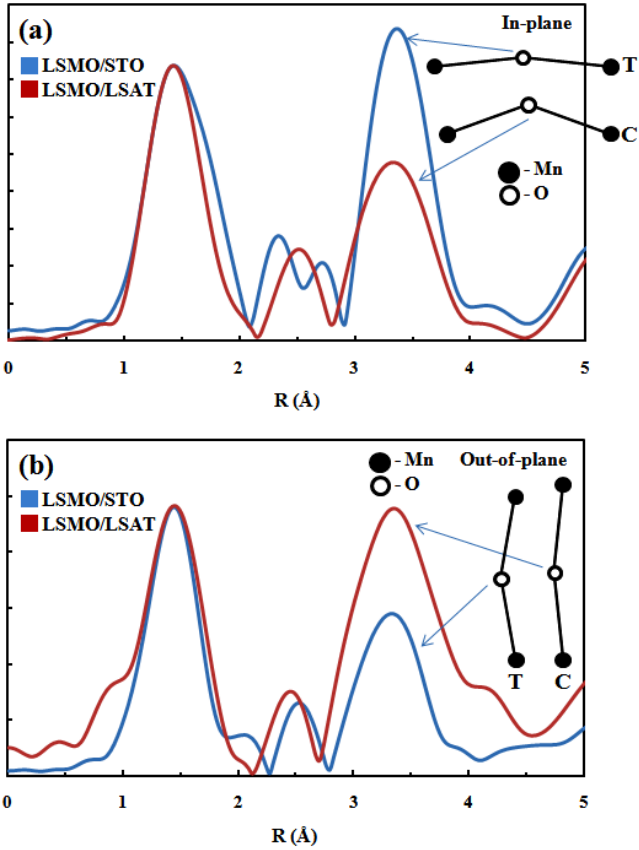

FIG. 7. The *R*-space Fourier transformed Mn K-edge EXAFS spectra of the LSMO thin films grown under compressive and tensile stresses on LSAT and STO substrates, respectively. Data are shown for the only in-plane bond contributions (a) and only for out-of-plane bond contributions (b). Letters C and T refer to Mn-O-Mn bond under compressive and tensile stresses, respectively.

## VII. Implications of $BO_6$ octahedra rotations on the magnetic properties

Due to a strong spin-orbit coupling, octahedral rotations are expected to influence some of the magnetic properties of SRO and LSMO materials. Number of studies have been reported on the magnetic anisotropy in SRO [33-37] and LSMO [38-44] thin films. Generally, reports of the uniaxial magnetic anisotropy in SRO thin films grown on STO(001) substrates strongly point to the importance of a mono-domain growth regime, even though the in-plane lattice of STO unit cell is square [34-37]. Mechanism of the mono-domain growth of SRO(110) thin films on STO(001) substrates was already established elsewhere [45]. While the uniaxial magnetic anisotropy in mono-domain SRO films grown on STO(001) has been linked to a crystalline anisotropy through octahedral rotations [34], the complete rotational pattern of the $RuO_6$ octahedra and the resulting symmetry of the SRO unit cell was not described. According to our study, the SRO unit cell grown on STO(001) substrate experience compressive stress and assume monoclinic [110] out-of-plane oriented unit cell with a space group $P2_1/m$. According to Glazer, the octahedral rotation pattern of such unit cell is not the same around $[100]_c$ and $[010]_c$ directions as shown in Figure 6(a) and (b). The difference breaks the symmetry along perpendicular in-plane directions and therefore might be the cause for the in-plane anisotropic magnetic properties in SRO films under compressive stress.

Single domain LSMO thin films grown on NGO(110), LSAT(110) substrates under compressive stress also exhibit in-plane uniaxial magnetic anisotropy [16,44]. The magnetic easy axis of a monoclinic LSMO(110) thin film grown on NGO(110) substrate was found to be along [1-10]$_o$ direction, while the hard axis is aligned along [001]$_o$ direction [44]. Besides a dissimilar in-plane MnO$_6$ octahedra rotation pattern, the uniaxial anisotropy in LSMO/NGO thin films can also be linked to the in-plane strain asymmetry arising from different in-plane lattice parameters of the NGO(110) substrate. The uniaxial magnetic anisotropy observed in LSMO films grown on LSAT(110) substrates cannot be attributed to the in-plane strain anisotropy since the in-plane lattice constants of a substrate are virtually the same [16]. It was shown that thin LSMO(110) films grown on LSAT(110) substrates exhibit uniaxial magnetic anisotropy with easy axis aligned along [001]$_o$ direction and hard axis along [1-10]$_o$ direction which can only be explained by a distinct MnO$_6$ octahedra rotation pattern around perpendicular in-plane directions of a monoclinic unit cell (space group P2$_1$/m) [16]. The uniaxial magnetic anisotropy was also observed in LSMO thin films grown on vicinal STO(001) substrates [41,46]. As it was shown above, the LSMO unit cell on STO(001) under tensile stress is tetragonal but with distinct octahedral rotations around perpendicular in-plane directions and no rotations around [001]$_c$ axis. The anisotropy in these films was reported to be step induced and it is not apparent how the in-plane asymmetry of octahedral rotations influence anisotropy in these films.

## VIII. Lattice modulations

We observed that LSMO thin films accommodate stress not only by $B$O$_6$ octahedra rotations. Under compressive stress thin films exhibit long range lattice modulations [47]. Figure 8 shows the reciprocal space maps of LSMO film grown on NGO taken around LSMO($hk$0) Bragg reflections with $h = k = 1, 2, 3, 4$. Besides the Bragg peaks, satellite peaks are clearly visible. The satellite peaks are aligned in-plane and their positions do not shift with the Bragg peak order indicating that the satellites are originating from the long range modulations which are periodic in-plane. Moreover, the satellites are only visible if the x-ray beam direction is parallel to the [001]$_o$ unit cell direction and are absent along [1-10]$_o$ direction demonstrating highly anisotropic nature of the long range modulations. Other groups also reported the existence of the satellite peaks in compressively strained LSMO thin films grown on LSAT substrates [48,49]. Zhou *et al*. attributed the appearance of the satellites to the recovery of the rhombohedral LSMO unit cell as thin film accommodates the mismatch stress by forming rhombohedral twins [48]. Jin *et al*. reported domains consisting of monoclinic LSMO unit cells which form two domain motifs that give rise to the satellites [49]. In both studies the rhombohedral and monoclinic unit cells were derived from the dissimilar positions of the asymmetric LSMO Bragg diffraction peaks along [100]$_c$, [010]$_c$, [-100]$_c$, and [0-10]$_c$ in-plane directions. Interestingly, they observed that the satellite peaks were positioned either higher or lower in Q$_z$ value with respect to the positions asymmetric LSMO{103} Bragg peaks. If we overlook $B$O$_6$ octahedra rotations, our monoclinic (110)-oriented unit cell shown in Figure 2 can be successfully described as a tilted pseudo-cubic (001)-oriented unit cell similar to the monoclinic one reported by Jin *et al*. [49]. However, in our case only (260)$_o$/(204)$_c$ and (620)$_o$/(-204)$_c$ peaks are at different positions while (444)$_o$/(024)$_c$ and (44-4)$_o$/(0-24)$_c$ peak positions are at the same positions. Such peak arrangement indicates that in our samples the pseudo-cubic unit cells are all tilted only along one direction, [1-10]$_o$/[010]$_c$, indicating the mono-domain state. The results unequivocally show that twin domains in this case are not responsible for the satellite peak formation. Moreover, the tilt direction of the pseudo-cubic unit cell does not agree with the direction of the long range modulations. In fact, they are 90 degree apart in-plane.

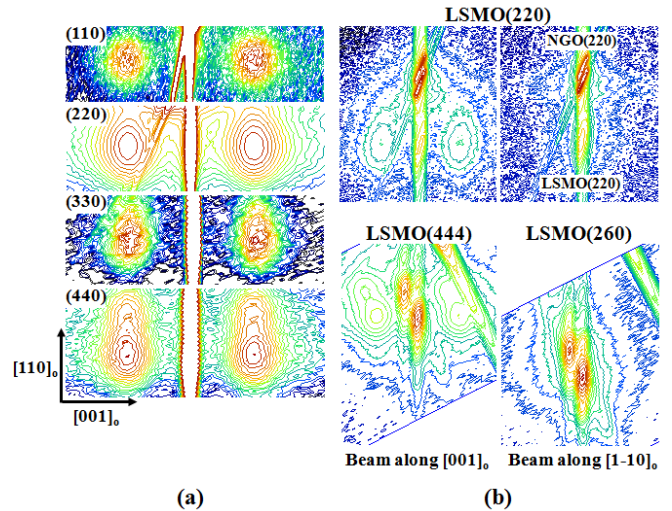

FIG. 8. Lattice modulations observed by XRD: (a) reciprocal lattice maps around LSMO ($hk$0) reflections with $h = k = 1,2,3,4$. (b) RLM's around LSMO (220), (444) and (620) reflections with incident beam aligned along the [001]$_o$ direction and along [1-10]$_o$ directions. Note that LSMO (444) and (260) peak positions are at different Q$_z$ values, same as in Fig. 2. The double peaks appear due to K$_{\alpha1}$ and K$_{\alpha2}$ wavelengths.

The observed results can be explained by assuming a monoclinic LSMO unit cell that is (110) out-of-plane oriented as shown in Figures 2 and 3. In bulk, besides the rhombohedral unit cell with the space group R-3c, the LSMO unit cell can be described as a monoclinic one with the symmetry group I2/c and $\beta$-angle slightly different from 90 degrees [23]. The variation of the $\beta$-angle in bulk unit cell alters the angle between *ab*-plane and the *c*-axis. In thin films, as it was shown above, the coherent growth and the lattice mismatch constrains the unit cell into a monoclinic P2$_1$/m space group where the $\gamma$-angle deviates from 90 degrees depending on the direction and the magnitude of strain. Under compressive stress, as shown in Figure 8(b), layers with such unit cell exhibit lattice modulations only along [001]$_o$ direction. In contrast to the observations by Zhou *et al*. and Jin *et al*., satellite peaks in our films are perfectly aligned in-plane with the asymmetric LSMO (444) and (44-4) Bragg peaks. No satellites were observed around LSMO (260) and (620) peaks which in-plane positions are different by 90 degrees from LSMO(444) and (44-4) peaks as shown in Figure 4. The origin of the modulations can be attributed to a deviation of an angle between *ab*-plane and the *c*-axis of the monoclinic unit cell. The modulations can then be represented as the unit cell displacements along the out-of-plane direction which are periodic in-plane along [001]$_o$ direction similar to those described elsewhere [50-53]. The existence of the satellite peaks along all [100]$_c$, [010]$_c$, [-100]$_c$, and [0-10]$_c$ in-plane directions and their misalignment along $Q_z$ with respect to the Bragg peak positions observed by Jin *et al*. unambiguously indicate a presence the 90 degree twinning of (110)-oriented monoclinic unit cells in LSMO films. It is interesting to note, that Pailloux *et al*. observed twins in LSMO/STO layers only for the substrates with a very low miscut angle. For thin films on substrates with 1.5° degree miscut angle twins were absent and only lattice modulations were visible. It was also reported that LSMO/STO thin films grown on vicinal substrates exhibit in-plane uniaxial magnetic anisotropy with the easy axis along the step edge direction [41,46]. The findings mentioned above point to the importance of the LSMO unit cell orientation which exhibits different octahedral rotation pattern along perpendicular in-plane directions to the magnetic properties of the LSMO thin films.

The observed lattice modulations can be quantitatively described using a kinematical x-ray diffraction. For simplicity, in the model we used a pseudo-cubic cubic LSMO unit cell with $a_{c\|} = c_o/2$ and $a_{c\perp} = [a_o^2 + b_o^2 - 2a_o b_o \cos(180 - \gamma_o)]/2$, where $a_{c\|}$ and $a_{c\perp}$ are the in-plane and out-of-plane pseudo-cubic lattice parameters, respectively. We assume that the unit cells have displacements along *L*-direction (out-of-plane) which are periodic along *H*-direction (in-plane) with periodicity, $\Lambda$. Then a one-dimensional complex structure factor along *H*-direction can be written as:

$$F_H = F_{uc} \sum_j e^{2\pi i (Hx_j + Lz_j)} =$$
$$F_{uc} \sum_{n=1}^{N} e^{2\pi i \left( H \frac{x_n}{a_{c\|}} + L \frac{z_n}{a_{c\perp}} \right)}, \qquad (5)$$

where $F_{uc}, x_j = \frac{x_n}{a_{c\|}}, z_j = \frac{z_n}{a_{c\perp}}$ are the unit cell structure factor and the relative x- and z-positions along *H* and *L* directions of a LSMO pseudo-cubic unit cell, respectively. *N* is the total number of the unit cells along *H*-direction. We consider that the unit cell displacements occur only along *L*-direction and they are periodic only along *H*-direction such that $x_n = na_{c\|}$ and $z_n = A_L \cos(k_H n a_{c\|})$, where *n* is a unit cell number, $A_L$ is modulation amplitude and $k_H = \frac{2\pi}{\Lambda}$ is a modulation wave vector. In this case the structure factor can be rewritten as:

$$F_H = F_{uc} \sum_{n=1}^{N} e^{2\pi i \left[ Hn + \frac{L}{a_{c\perp}} A_L \cos(k_H n a_{c\|}) \right]}. \qquad (6)$$

In order to account for the satellite peak broadening, we assume that the modulation period deviates from a mean value $\Lambda$ according to a Gaussian distribution with a standard deviation, $\delta\Lambda$. The simulation results together with experimental data for LSMO/LSAT and LSMO/NGO samples are shown in Figure 9. As can be seen from the Figure, the kinematical model was able to reproduce all features of the XRD spectra: main Bragg peak, satellite peak positions, intensities and widths using single parameters: for LSMO/LSAT $\Lambda$ = 30 nm, $A_L$ = 0.40 Å and $\delta\Lambda$ = 5 nm, and for LSMO/NGO $\Lambda$ = 23 nm, $A_L$ = 0.18 Å and $\delta\Lambda$ = 5 nm. The simulations were also performed assuming unit cell displacements along the in-plane $[1-10]_o$ direction. However, it did not produce satisfactory results indicating that unit cell displacements occur only along out-of-plane direction which is consistent with the coherent layer growth on a single crystal substrate. The highly anisotropic nature of the lattice modulations further confirms the statement that the stress is accommodated differently along perpendicular in-plane directions in LSMO and SRO thin films.

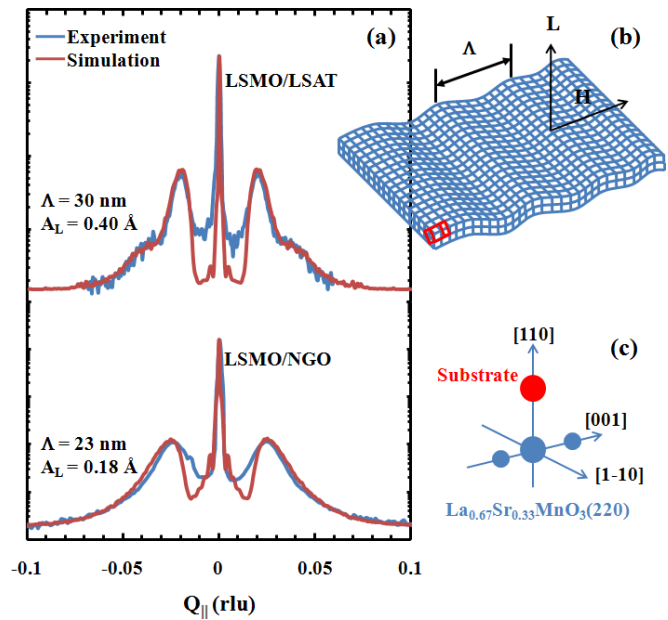

Fig. 9. Lattice modulations: (a) observed (blue) and simulated (red) XRD profiles around LSMO(220) reflections for films grown on LSAT and NGO substrates. (b) schematic drawing of the lattice modulations used in calculations; (c) schematic picture of a reciprocal space showing substrate peak (red) together with LSMO(220) peak and its first order satellites (blue). Here we used $Q_\parallel = 4\pi \sin(\theta/\lambda)$, where $\theta$ is the Bragg angle and $\lambda$ =1.540598 Å.

## IX. Conclusions

We have demonstrated that under epitaxial strain thin LSMO and SRO films behave very similarly: under compressive stress they have (110) out-of-plane oriented monoclinic unit cell with space group $P2_1/m$ (No. 11), while under tensile stress both films exhibit [001] out-of-plane oriented tetragonal unit cell with space group Cmcm (No. 63). In both cases $BO_6$ octahedra are rotated in-phase around $[100]_c$ direction and out-of-phase around $[010]_c$ direction. Such rotational pattern might affect some physical properties in these materials such as in-plane magnetic anisotropy in thin LSMO films. Besides the in-plane rotations, $BO_6$ octahedra under compressive strain are also rotated out-of-phase around $[001]_c$ direction while such rotations are absent under tensile strain. The additional strain along $[001]_o$ direction is accommodated by periodic lattice modulations. We believe, that the observed lattice response to the epitaxial strain is of a general nature and can be applied to other perovskite-like materials which possess bulk orthorhombic or rhombohedral structures.

## Acknowledgments

This work is supported in part by the Department of Energy, Office of Basic Energy Sciences, Division of Materials Sciences and Engineering, under contract DE-AC02-76SF00515. This research was financially supported by the Dutch Science Foundation and by NanoNed, a nanotechnology program of the Dutch Ministry of Economic Affairs.